\documentclass[conference]{IEEEtran}
\IEEEoverridecommandlockouts

\usepackage{cite}
\usepackage{amsmath,amssymb,amsfonts}
\usepackage{algorithmic}
\usepackage{graphicx}
\usepackage{textcomp}
\usepackage{xcolor}
\usepackage{multirow}
\usepackage{url}
\usepackage{hyperref}
\usepackage{subcaption}
\def\BibTeX{{\rm B\kern-.05em{\sc i\kern-.025em b}\kern-.08em
    T\kern-.1667em\lower.7ex\hbox{E}\kern-.125emX}}
\begin{document}

\title{H-SAGE: Holistic Speaker-Aware Guided Experts for MoE-based Multi-Talker ASR \\
\thanks{$^{*}$ Corresponding author. This work was supported by the National Natural
 Science Foundation of China (62271270).}
}

\author{\IEEEauthorblockN{Yujie Guo$^{1}$, Jiaming Zhou$^{1}$, Yuhang Jia$^{1}$, Yang Chen$^{1}$, and Yong Qin$^{1,*}$}
\IEEEauthorblockA{
    $^{1}$TMCC, College of Computer Science, Nankai University, Tianjin, China\\
    \{guoyujie02,zhoujiaming,2120240729,2120230617\}@mail.nankai.edu.cn, qinyong@nankai.edu.cn
    }
}
\maketitle

\begin{abstract}
Multi-talker Automatic Speech Recognition (MTASR) faces significant challenges in accurately transcribing overlapping speech, particularly under complex high-overlap conditions. While recent Mixture-of-Experts (MoE) approaches have shown promise, they typically rely on frame-independent routing that leads to temporal myopia, and depend solely on the downstream ASR objective, which results in implicit and ungrounded representation learning. To address these limitations, we propose Holistic Speaker-Aware Guided Experts (H-SAGE) for MoE-based MTASR. Specifically, we introduce a Speaker-Aware Global Encoder to capture long-term dependencies, supervised by an auxiliary Overlap-Aware Loss that explicitly guides the model to discern acoustic states. Furthermore, we design a Holistic Gating Mechanism to arbitrate expert selection by jointly evaluating global context and local details. Experiments on LibriSpeechMix demonstrate that H-SAGE achieves consistent improvements over strong baselines, particularly in complex scenarios, validating that explicit acoustic guidance effectively enhances expert collaboration. Our code can be found at \url{https://github.com/NKU-HLT/H-SAGE}.
\end{abstract}

\begin{IEEEkeywords}
multi-talker automatic speech recognition, cocktail party problem, Mixture-of-Experts, speaker-aware
\end{IEEEkeywords}

\section{Introduction}
In recent years, Automatic Speech Recognition (ASR) has witnessed remarkable progress due to the advent of deep learning. State-of-the-art architectures have achieved human-level performance in clean, single-speaker scenarios. However, these systems often degrade significantly in realistic conversational environments, which are characterized by spontaneous turn-taking and frequent speech overlaps. This limitation has catalyzed the emergence of Multi-Talker ASR (MTASR) as a critical research frontier. Unlike traditional ASR, which assumes a monotonic mapping between a single acoustic stream and text, MTASR aims to simultaneously disentangle and transcribe overlapping utterances from multiple speakers, addressing the classic ``Cocktail Party Problem''.

Current end-to-end MTASR approaches fall into two paradigms: Single-Input Multiple-Output (SIMO) and Single-Input Single-Output (SISO)~\cite{survey}. SIMO-based methods~\cite{SIMO1,SIMO2, SIMO3} utilize Permutation Invariant Training (PIT)~\cite{PIT1} to assign speech to specific branches. While effective, they are constrained by a fixed number of output branches, limiting their flexibility when the speaker count is unknown or variable.

In contrast, SISO approaches employ Serialized Output Training (SOT)~\cite{SOT} to generate a serialized transcription containing all speakers. This paradigm inherently handles variable speaker counts, offering superior flexibility. However, it relies heavily on the attention mechanism to implicitly disentangle overlapping speech. Without explicit guidance, the model often struggles to distinguish speakers in high-overlap segments, resulting in performance degradation.

To mitigate this issue, recent research has pivoted towards augmenting SISO frameworks with explicit speaker modeling, utilizing auxiliary modules or objectives to capture discriminative speaker characteristics. For instance, CSE-SOT~\cite{cse-sot} employs multi-branch structures to capture distinct speaker representation. Similarly, SACTC~\cite{sactc} and SD-CTC~\cite{sdctc} leverage improved connectionist temporal classification to explicitly guide the encoders in separating speaker information. Furthermore, SSA~\cite{SSA} utilizes an external diarization system to explicitly provide speaker activity cues, guiding the ASR model to generate speaker-specific transcriptions.

Parallel to these advancements, the Mixture-of-Experts (MoE) paradigm has recently been explored in MTASR, with GLAD~\cite{glad} representing an early attempt. Despite integrating global and local routers, GLAD suffers from key structural limitations. Its global information relies on frame-independent projections, preventing effective modeling of speaker-level temporal dynamics. Moreover, its speaker-aware behavior is learned only implicitly through the standard ASR training objective, making the router prone to shortcut learning rather than acquiring robust discriminative cues. Finally, its expert selection strategy is derived solely from local representations, giving the router a narrow, non-holistic perspective that limits its adaptability to complex acoustic conditions.

To address these issues, we propose \textbf{H}olistic \textbf{S}peaker-\textbf{A}ware \textbf{G}uided \textbf{E}xperts (H-SAGE), a framework that rethinks expert routing for MoE-based MTASR. H-SAGE comprises two complementary components: a Speaker-Aware Global Encoder (SA-Encoder) that explicitly models speaker activity states—such as silence, single-speaker, and overlap—guided by an auxiliary Overlap-Aware Loss ($\mathcal{L}_{OA}$), and a Holistic Gating Mechanism that assigns experts by jointly leveraging global acoustic context and local frame-level cues. Together, these components deliver more reliable and context-informed routing decisions, addressing the core limitations of prior MoE-based approaches. Our contributions are as follows:

\begin{figure*}[t]
  \centering
  \includegraphics[width=0.9\textwidth]{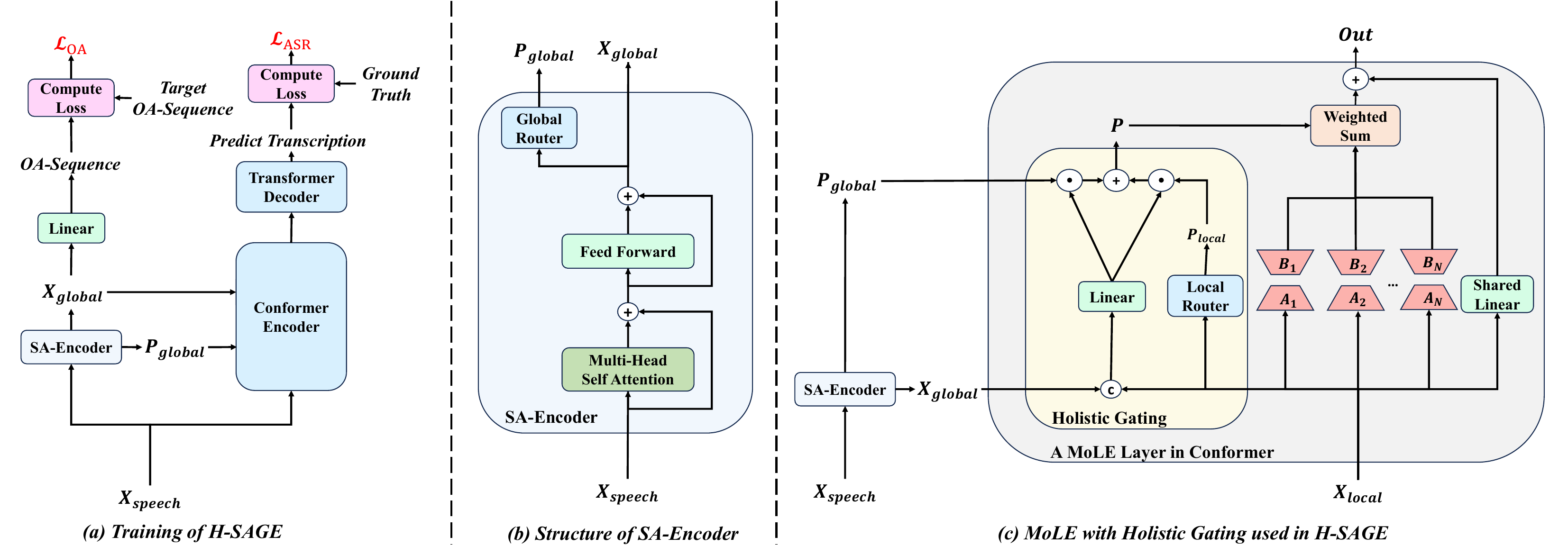}
  \caption{Overview of the H-SAGE architecture. (a) Training Pipeline: The Conformer Encoder integrates the MoLE blocks shown in (c). The model is jointly optimized by $\mathcal{L}_{ASR}$ and the explicit Overlap-Aware Loss ($\mathcal{L}_{OA}$). (b) Structure of SA-Encoder: SA-Encoder extracts global context from the convolutional frontend output $X_{speech}$, producing global routing probabilities $P_{global}$ and context features $X_{global}$. (c) MoLE used in H-SAGE: Detailed structure of the MoLE block. These blocks replace all the linear transformations in the Conformer Encoder. The Holistic Gating mechanism fuses global context ($X_{global}$) and local input ($X_{local}$) to adaptively balance routing probabilities for precise expert selection.}
  \label{fig:h-sage}
  \vspace{-4mm}
\end{figure*}

\begin{itemize}
    \item We propose H-SAGE, a novel framework anchored by a SA-Encoder. Supervised by an auxiliary Overlap-Aware Loss, this design captures evolving acoustic states and shifts the routing paradigm from implicit representation learning to explicit acoustic modeling.
    \item We introduce a Holistic Gating Mechanism to optimize expert assignment. By synthesizing global acoustic contexts with local representations, it ensures a comprehensive perspective for robust expert selection. 
    \item Experiments on LibriSpeechMix demonstrate consistent improvements over strong baselines in multi-talker scenarios. Crucially, further analysis validates the effectiveness of the proposed mechanisms, confirming their role in expert collaboration and robust speaker disentanglement.
\end{itemize}

\section{Preliminary Knowledge}
\subsection{Serialized Output Training}

Serialized Output Training (SOT)~\cite{SOT} serves as a prominent Single-Input Single-Output (SISO) paradigm in MTASR. In contrast to architectures relying on multiple output branches, SOT leverages a single decoder to transcribe overlapping speech from a variable number of speakers into a unified sequence. Given an utterance with $N$ speakers, the model is trained to generate a concatenated sequence formatted as $Y = \{text_1, \langle sc \rangle, text_2, \dots, \langle sc \rangle, text_N\}$, where $text_i$ represents the transcription sequence of the $i$-th speaker, and $\langle sc \rangle$ denotes a special speaker-change token. We adopt the First-In, First-Out (FIFO) strategy, organizing transcripts chronologically based on speaker onset. This serialization strategy allows the model to implicitly disentangle overlapping speech, effectively bypassing the combinatorial complexity associated with permutation-based methods.

\subsection{MoE-based MTASR}

GLAD~\cite{glad} represents a pioneering effort in applying the Mixture-of-Low-Rank-Experts (MoLE) paradigm to MTASR. It replaces all linear layers within the encoder with MoLE blocks, whose structure resembles MoE-LoRA~\cite{mola}. Given an input feature $X \in \mathbb{R}^{d_{in}}$, the output is formulated as:

\vspace{-2mm} 
\begin{equation}
    Y = W X 
    + \frac{\alpha}{r}  {\textstyle \sum_{i=1}^{N}}  P_i B_i A_i X
    + b,
    \label{eq:mole}
\end{equation}
\vspace{-3mm} 

\noindent where $W$ and $b$ denote the weights and bias of the shared linear layer, respectively. Adopting the structural design of LoRA~\cite{lora}, each Low-Rank expert is parameterized by low-rank matrices $A_i \in \mathbb{R}^{r \times d_{in}}$ and $B_i \in \mathbb{R}^{d_{out} \times r}$ with rank $r \ll \min(d_{in}, d_{out})$, scaled by a factor $\alpha$. The expert weight $P_i$ is computed via a gating mechanism that combines global and local routing signals.

The expert activation weight $P_i$ is computed via a gating mechanism that combines global and local routing signals. Specifically, GLAD employs a linear projection on the convolutional frontend output to derive a frame-independent global router, while the local router is derived from the current layer's hidden states. The fusion coefficient is derived from the current layer's features to balance these signals. Consequently, the final routing weight fuses the frame-independent global projection with instantaneous local representations.

\section{Our Methods}
As illustrated in Fig.~\ref{fig:h-sage} (a), H-SAGE is built upon an MoE-based encoder-decoder architecture. Consistent with prior work~\cite{glad}, we replace standard linear layers in the encoder with MoLE blocks. Uniquely, our architecture integrates two key modules: the SA-Encoder module (Fig.~\ref{fig:h-sage} (b)), which captures global acoustic context supervised by the Overlap-Aware Loss, and the Holistic Gating Mechanism (Fig.~\ref{fig:h-sage} (c)), which synthesizes global acoustic contexts with local representations for expert routing.

\subsection{Speaker-Aware Global Encoder with Explicit Supervision}

\begin{figure}[t]
  \centering
  \includegraphics[width=0.45\textwidth]{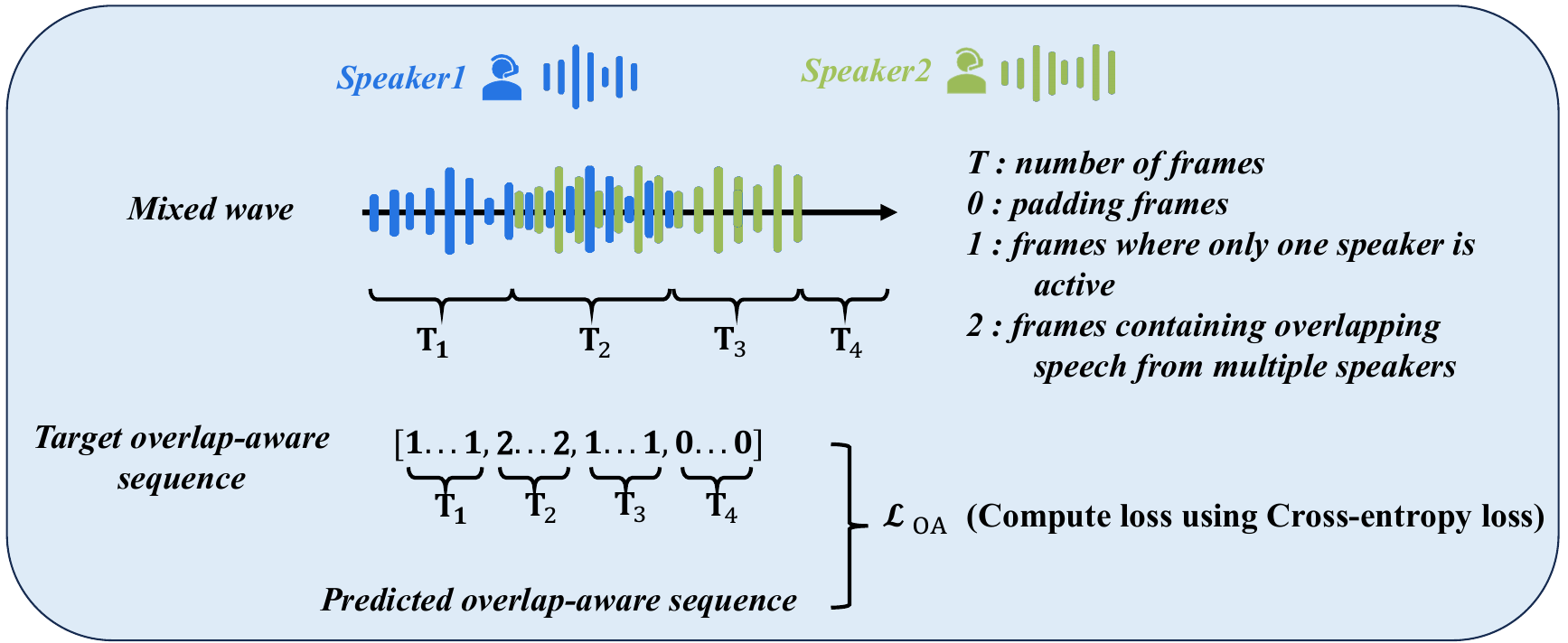}
  \caption{The detail of Overlap-Aware Loss}
  \label{fig:oaloss}
  \vspace{-4mm}
\end{figure}

To transition from implicit adaptation to explicit acoustic modeling, we introduce SA-Encoder. Unlike frame-independent projections that suffer from temporal myopia, SA-Encoder is designed to capture dynamic speaker activity states via long-term dependency modeling, explicitly supervised by an auxiliary Overlap-Aware Loss ($\mathcal{L}_{OA}$).

\textbf{SA-Encoder}: As illustrated in Fig.~\ref{fig:h-sage} (b), SA-Encoder operates directly on the signals $X_{speech} \in \mathbb{R}^{T \times D}$ extracted by the convolutional frontend. This design effectively preserves rich acoustic characteristics that are often diluted in deeper network layers. Deviating from simple linear projection, SA-Encoder incorporates a Multi-Head Self-Attention block followed by a Feed-Forward Network. This architecture enables the model to capture long-range temporal dependencies, which is pivotal for differentiating between transient noise and sustained speech overlaps. Through this process, SA-Encoder extracts the global context representation $X_{global} \in \mathbb{R}^{T \times D}$. Subsequently, $X_{global}$ is transformed via the global router, which applies a linear projection and a softmax function to generate the global routing probabilities $P_{global}$, which serve as a critical input for the holistic gating mechanism.

\textbf{Explicit Supervision via Overlap-Aware Loss}: Although SA-Encoder extracts global context, without explicit guidance, the learned representations $X_{global}$ remain semantically ungrounded. The model might rely on shortcut features, settling into suboptimal local minima. To enforce explicit acoustic modeling, we introduce the Overlap-Aware Loss ($\mathcal{L}_{OA}$).

As illustrated in Fig.~\ref{fig:h-sage} (a), we utilize $X_{global}$ to predict a frame-level overlap-aware sequence. As visualized in Fig.~\ref{fig:oaloss}, we define the target sequence $Y_{overlap\_aware} = \{y_1, y_2,\dots, y_T\}$, where each token $y_t \in \{0, 1, 2\}$ represents the current speaker state:
\begin{itemize}
\item 0 (Padding): Represents padding frames during training.
\item 1 (Single-Speaker): Represents frames where only one speaker is active.
\item 2 (Overlap): Represents frames where speech from multiple speakers overlaps.
\end{itemize}

The target sequence $Y_{overlap\_aware}$ is automatically generated from the temporal boundaries determined by each source utterance’s delay and duration. The delay is the offset time at which the utterance is inserted into the mixture, while the duration reflects the length of the original clean segment; together, they define the active time span of each speaker in the mixed signal.

Correspondingly, the predicted overlap-aware sequence is obtained by projecting the global context representation $X_{global}$ through a classification head. The Overlap-Aware Loss $\mathcal{L}_{OA}$ is then computed as the Cross-Entropy between the predictions and the targets $Y_{overlap\_aware}$. Minimizing this objective explicitly supervises the model to distinguish between single-talker and multi-talker segments, providing robust guidance for expert selection.

\subsection{Holistic Gating Mechanism}

Even with the robust acoustic states captured in $P_{global}$, relying solely on local features to determine the fusion ratio, as done in previous works, creates an information asymmetry. This restricted perspective prevents the model from optimally balancing global context with local details. To address this, we introduce the Holistic Gating Mechanism, which arbitrates expert selection by jointly evaluating both information sources.

Specifically, given the local features $X_{local} \in \mathbb{R}^{T \times D_1}$ and the global context $X_{global} \in \mathbb{R}^{T \times D_2}$ extracted from SA-Encoder, we concatenate them to form a holistic representation. As illustrated in Fig.~\ref{fig:h-sage} (c), the fusion weights $\beta \in \mathbb{R}^{T \times 2}$ are computed as:

\vspace{-1mm} 
\begin{equation}
\begin{aligned}
    X_{holistic} = Concatenate(X_{local}, X_{global}), \\
    \beta = softmax(X_{holistic} W_{HG} + B_{HG}),
\end{aligned}
\label{eq:sage}
\end{equation}
\vspace{-2mm} 

\noindent where $X_{holistic} \in \mathbb{R}^{T \times (D_1+D_2)}$ represents the unified feature representation. $W_{HG} \in \mathbb{R}^{(D_1 + D_2) \times 2}$ and $B_{HG} \in \mathbb{R}^2$ denote the weight matrix and bias vector of the linear transformation, respectively.

To generate the final expert activation probabilities, we utilize $\beta$ to dynamically weight the local routing distribution $P_{local} \in \mathbb{R}^{T \times N}$ and the global routing distribution $P_{global} \in \mathbb{R}^{T \times N}$, where $N$ is the number of experts. Let $\beta_{0}, \beta_{1} \in \mathbb{R}^{T \times 1}$ denote the decomposed weights for the local and global components derived from $\beta$, respectively. The final probability for the $i$-th expert, $P_{i} \in \mathbb{R}^{T \times 1}$, is obtained by:

\vspace{-1mm} 
\begin{equation}
\begin{aligned}
    P_{i} = \beta_{0} \odot P_{local,i} + \beta_{1} \odot P_{global.i}
\end{aligned}
\label{eq:expet weight}
\end{equation}
\vspace{-4mm} 

\noindent where $\odot$ denotes element-wise multiplication.

Once the final expert weights $P$ are generated via this holistic view, the output of the MoLE layer is computed following Eq.~\ref{eq:mole}. By adaptively integrating SA-Encoder-extracted speaker cues with local phonetic details, the proposed mechanism ensures that expert usage is guided by a comprehensive perspective, thereby enhancing the model's ability to disentangle speaker identity and transcribe spoken content.

\subsection{Multi-Task Training Objective}

To jointly optimize the model for both accurate transcription and explicit acoustic cognition, we employ a multi-task learning strategy. The total objective function $\mathcal{L}$ is formulated as a weighted sum of the primary recognition loss and the auxiliary Overlap-Aware loss:

\vspace{-1mm} 
\begin{equation}
\begin{aligned}
    \mathcal{L} = \mathcal{L}_{ASR} + \lambda \mathcal{L}_{OA}, 
\end{aligned}
\label{eq:loss}
\end{equation}
\vspace{-4mm} 

\noindent where $\mathcal{L}_{ASR}$ denotes the standard cross-entropy loss calculated on the serialized output targets for the MTASR task. $\mathcal{L}_{OA}$ is the auxiliary Overlap-Aware Loss, which enforces the learning of discriminative acoustic states. The hyperparameter $\lambda$ controls the weight of the explicit acoustic supervision, balancing the trade-off between speaker disentanglement and transcription performance.

\section{Experimental Setup}

\begin{table}[t]
\renewcommand{\arraystretch}{1.2}
\centering
\caption{Training dataset composition}
\label{tab:traindata}
\begin{tabular}{c|c|ccc|c}
\hline
\multirow{2}{*}{} & \multirow{2}{*}{single-speaker} & \multicolumn{3}{c|}{2-speaker} & \multirow{2}{*}{Total} \\ \cline{3-5}
                  &                                 & Low      & Mid      & High     &                        \\ \hline
Utt.              & 202493                          & 81808    & 75245    & 45423    & 404985                 \\
Dur.(hrs)         & 692.2                           & 491.8    & 383.0    & 203.2    & 1770.2                 \\ \hline
\end{tabular}%
\vspace{-3mm}
\end{table}

\subsection{Dataset}
Our experiments are conducted on the LibriSpeechMix (LSM) benchmark~\cite{SOT}, which includes both 2-speaker (LSM-2mix) and 3-speaker (LSM-3mix) scenarios derived from LibriSpeech~\cite{Librispeech}. Given that the original LSM lacks a training partition, we construct a custom training set following the protocol in~\cite{cse-sot, sactc, glad}. Specifically, we synthesize 2-speaker mixtures by randomly pairing LibriSpeech utterances with varying temporal offsets. To construct the final training corpus, we sample a subset of these simulated mixtures and combine them with a portion of the original single-speaker utterances.

To facilitate a fine-grained evaluation across multi-talker scenarios of varying complexity, we stratify the dataset based on the overlap rate, defined as the duration of the overlapping segment divided by the total duration of the mixed speech. We categorize the data into three levels: Low $(0, 0.2]$, Medium $(0.2, 0.5]$, and High $(0.5, 1.0]$. This division enables a systematic analysis of model robustness under different acoustic conditions. The detailed statistics and distribution of the resulting dataset are summarized in Table~\ref{tab:traindata}.

\subsection{Model Configuration and Training Setup}
All models are implemented using the ESPnet2 toolkit~\cite{espnet}. The architecture follows an encoder-decoder framework, featuring a Conformer encoder~\cite{conformer} and a 6-block Transformer decoder. Regarding the internal structure, both encoder and decoder blocks utilize 4-head self-attention with 256 hidden units. In particular, the encoder employs macaron-style blocks equipped with 1024-dimensional feed-forward modules, whereas the decoder blocks incorporate 2048-dimensional feed-forward layers.

In our experiments, SOT~\cite{SOT}, SOT-SACTC~\cite{sactc}, and GLAD-SOT~\cite{glad} serve as the baseline systems. Additionally, to evaluate the efficacy of MoE without global guidance, we implement an SOT + Local MoLE baseline, which incorporates local Low-Rank experts but relies solely on local routing.

To eliminate the influence of model size, we control the number of encoder layers for each model to ensure parameter parity. Accordingly, we assign 14 encoder layers to SOT, 13 to SOT-SACTC, and 12 to SOT + Local MoLE, GLAD-SOT, and our H-SAGE. As detailed in Table~\ref{tab:res}, this setup guarantees that all compared models possess a similar number of parameters.

For baseline systems, we adopt the optimal hyperparameters reported in their respective papers. For H-SAGE, the MoLE module is configured with 3 experts, setting both the rank $r$ and scaling factor $\alpha$ in Eq.~\ref{eq:mole} to 8, while the OA-Loss coefficient $\lambda$ in Eq.~\ref{eq:loss} is set to 3. All models are trained for 35 epochs. Training is performed on 8 NVIDIA GeForce RTX 3090 GPUs using the Adam optimizer with a peak learning rate of $5e^{-4}$ and 25,000 warm-up steps.

\subsection{Evaluation Metrics}

Consistent with prior studies~\cite{cse-sot, sactc, glad}, we evaluate single-talker performance using Word Error Rate (WER). For multi-talker scenarios, we employ Permutation Invariant WER (PI-WER)~\cite{SOT} to resolve speaker permutation ambiguity. Furthermore, to provide a fine-grained assessment of robustness, we report Overlap-Aware WER (OA-WER), which is calculated by averaging the WERs across the stratified overlap levels defined previously.

It is worth noting that while all models are trained exclusively on single-speaker and 2-speaker data, our evaluation extends to 3-speaker mixtures. Specifically, we use the single- and 2-speaker test sets to assess performance on in-domain scenarios, whereas the 3-speaker test set is utilized to evaluate the model's zero-shot generalization capability to unseen speaker numbers.

\section{Results and Discussion}
In this section, we validate H-SAGE, which demonstrates superior performance over existing approaches (Tables~\ref{tab:res} and~\ref{tab:res2}). We provide a detailed analysis covering performance comparisons (Table~\ref{tab:res}), followed by component-wise ablation studies (Table~\ref{tab:res2}) verifying the critical role of the proposed explicit modeling and holistic routing mechanisms. Finally, we examine the sensitivity of the auxiliary loss weight $\lambda$ to determine the optimal balance for explicit supervision.

\begin{table*}[t]
\caption{Main performance comparison (WER \%) against state-of-the-art baselines on LibriSpeech and LibriSpeechMix benchmarks. Consistent with standard protocols, all models are trained on a unified dataset comprising single-talker and two-talker utterances. The best and second-best results are highlighted in bold underline and underline, respectively.}
\label{tab:res}
\renewcommand{\arraystretch}{1.25}
\centering
\setcounter{table}{1}
\resizebox{\textwidth}{!}{%
\begin{tabular}{c|l|c|cc|cccccc|cccccc}
\hline
\multirow{3}{*}{System} & \multicolumn{1}{c|}{\multirow{3}{*}{Method}} & \multirow{3}{*}{Parm.(M)} & \multicolumn{2}{c|}{Librispeech}             & \multicolumn{6}{c|}{LSM-2mix}                                                                     & \multicolumn{6}{c}{LSM-3mix (zero-shot generalization)}                                                                            \\ \cline{4-17} 
                        & \multicolumn{1}{c|}{}                        &                           & \multirow{2}{*}{Dev} & \multirow{2}{*}{Test} & \multicolumn{2}{c|}{Overall}                     & \multicolumn{4}{c|}{Test(Conditional)}                    & \multicolumn{2}{c|}{Overall}                       & \multicolumn{4}{c}{Test(Conditional)}                         \\ \cline{6-17} 
                        & \multicolumn{1}{c|}{}                        &                           &                      &                       & Dev          & \multicolumn{1}{c|}{Test}         & low          & mid          & high         & OA-WER       & Dev           & \multicolumn{1}{c|}{Test}          & low           & mid           & high          & OA-WER        \\ \hline
S1                      & SOT~\cite{SOT}                                          & 36.07                     & 4.1                  & 4.5                   & 8.5          & \multicolumn{1}{c|}{8.3}          & 6.2          & 8.9          & 12.8         & 9.3          & 23.8          & \multicolumn{1}{c|}{24.2}          & 17.1          & 23.8          & 32.3          & 24.4          \\
S2                      & SOT + Local MoLE                                   & 35.09                     & 3.7                  & \textbf{3.8}          & \underline{6.4}    & \multicolumn{1}{c|}{6.5}          & 5.3          & 6.3          & 10.3         & 7.3          & 21.1          & \multicolumn{1}{c|}{21.7}          & 15.3          & 21.6          & 28.5          & 21.8          \\
S3                      & SOT-SACTC~\cite{sactc}                                    & 35.77                     & \underline{3.6}            & \textbf{3.8}          & 6.9          & \multicolumn{1}{c|}{6.7}          & \underline{4.8}    & 7.2          & 10.8         & 7.6          & 20.6          & \multicolumn{1}{c|}{20.0}          & \textbf{12.8} & 20.0          & 27.2          & \underline{20.0}    \\
S4                      & GLAD-SOT~\cite{glad}                                         & 35.18                     & \textbf{3.5}         & \underline{3.9}             & \textbf{6.0} & \multicolumn{1}{c|}{\underline{6.2}}    & 5.1          & \underline{6.4}    & \underline{8.9}    & \underline{6.8}    & \underline{19.9}    & \multicolumn{1}{c|}{\underline{19.8}}    & \underline{15.0}    & \underline{19.5}    & \underline{25.5}    & \underline{20.0}    \\
S5                      & H-SAGE                                       & 35.75                     & \underline{3.6}            & \textbf{3.8}          & \textbf{6.0} & \multicolumn{1}{c|}{\textbf{5.7}} & \textbf{4.7} & \textbf{5.7} & \textbf{8.2} & \textbf{6.2} & \textbf{19.7} & \multicolumn{1}{c|}{\textbf{19.5}} & 15.5          & \textbf{18.9} & \textbf{24.9} & \textbf{19.8} \\ \hline
\end{tabular}%
}
\vspace{-3mm}
\end{table*}

\subsection{Main Results}
In Table~\ref{tab:res}, we benchmark the proposed H-SAGE (S5) against representative baselines, analyzing the performance evolution across different architectural paradigms.

\textbf{Performance Evolution of Baselines}: 
First, comparing the standard SISO baseline SOT (S1) with SOT + Local MoLE (S2), we observe that S2 achieves consistent improvements despite having fewer parameters. This confirms that the effectiveness of S2 stems from the efficient decomposition of diverse acoustic patterns via specialized local experts. 
Moreover, SOT-SACTC~\cite{sactc} (S3) improves performance by introducing an explicit separation constraint. By utilizing speaker-aware CTC, S3 guides the encoder to learn disentangled speaker representations in the latent space, proving particularly effective in low-overlap scenarios. 
Meanwhile, GLAD-SOT~\cite{glad} (S4) advances the MoE paradigm of S2. Instead of relying on static local experts, S4 employs a global-local routing mechanism. The performance of S4 demonstrates that for MoE architectures, dynamic, context-aware expert selection is as critical as the experts themselves.

\textbf{Superiority of H-SAGE}: 
Building upon these advancements, H-SAGE (S5) addresses the structural limitations of GLAD and achieves the best overall performance. On the single-speaker LibriSpeech benchmark, H-SAGE maintains parity with strong baselines, confirming that the model incurs no degradation in clean scenarios. In the two-speaker mixtures (LSM-2mix), S5 consistently surpasses all baselines, achieving notable improvements in the challenging mid and high overlap conditions. These results suggest that by explicitly modeling overlap states, H-SAGE effectively resolves complex feature entanglement that hinders previous methods.

\textbf{Zero-Shot Generalization (LSM-3mix)}: 
The advantages of H-SAGE are most pronounced in the zero-shot LSM-3mix scenario, which evaluates generalization to unseen speaker counts. While baselines struggle to maintain performance, H-SAGE maintains a significant lead overall, particularly in complex high-overlap scenarios. 
We note a performance gap in the LSM-3mix-Low compared to SACTC. This is intuitive: SACTC's explicit separation constraints are highly efficient for sparse, easy-to-separate segments. In contrast, H-SAGE is optimized for complex entanglement. Nevertheless, H-SAGE's dominance in the more realistic mid and high overlap scenarios validates its superior robustness and generalization capability where it matters most.

\subsection{Ablation Study}

As summarized in Tables~\ref{tab:res2}, we  conduct an ablation study (S5--S8) to validate the contributions of core components.

\textbf{Effectiveness of Explicit Acoustic Supervision}: We compare the full model (S5) against the variant without explicit supervision (S6) to validate the role of $\mathcal{L}_{OA}$. On the clean Librispeech dataset, S5 exhibits a marginal performance drop compared to S6. This can be attributed to the interference from the auxiliary objective. For clean speech, explicit overlap detection is redundant. Consequently, the auxiliary task introduces optimization noise that marginally degrades S5 compared to S6 in the single-speaker scenario. 

However, in the in-domain LSM-2mix scenario, S5 surpasses S6. This indicates that explicit supervision effectively sharpens the router's decision boundaries, enabling more precise expert specialization than implicit learning alone. Most notably, S5 achieves substantial gains over S6 in the zero-shot LSM-3mix scenario. This result highlights the critical value of $\mathcal{L}_{OA}$ for generalization. In contrast to S6, which tends to overfit to two-speaker patterns, S5 leverages explicit supervision to capture a generic overlap state. This enables robust generalization to unseen scenarios with more speakers, demonstrating the vital role of explicit supervision.

\textbf{Necessity of Long-Range Temporal Modeling}: To isolate the structural contribution of the global encoder, we compare S6 (H-SAGE w/o OA-Loss $\mathcal{L}_{OA}$) against S7 (H-SAGE w/o SA-Encoder). It is important to note that in S7, the absence of the SA-Encoder implies that the global encoder degenerates to the simple frame-independent linear projection used in GLAD. 

The results show that S6 yields consistent improvements over S7 across the majority of metrics. This indicates that even without explicit supervision, the self-attention structure of SA-Encoder inherently captures long-range temporal dependencies inaccessible to the linear projection in S7. Unlike the instantaneous mapping in S7, SA-Encoder aggregates information across the sequence, proving that modeling temporal evolution of speech is a prerequisite for robust expert assignment.

\textbf{Necessity of Holistic Gating Mechanism}: We compare H-SAGE (S5) with the variant excluding holistic gating (S8). Note that in S8, the router degenerates to the GLAD fusion mechanism, relying solely on the current layer's features. We observe comparable performance in the single-speaker scenario. This suggests that in clean acoustic environments, local features provide sufficient guidance. However, H-SAGE consistently outperforms S8 across all other benchmarks, including in-domain LSM-2mix and zero-shot LSM-3mix scenarios. This confirms that while local cues suffice for simple tasks, the holistic view is indispensable for resolving complex feature entanglement in challenging multi-talker environments.

\begin{table}[t]
\setcounter{table}{2}
\caption{Ablation study and architectural analysis on the LibriSpeechMix benchmark.}
\label{tab:res2}
\renewcommand{\arraystretch}{1.3}
\centering
\resizebox{0.49 \textwidth}{!}{%
\begin{tabular}{c|l|cc|cc|cc}
\hline
\multirow{2}{*}{System} & \multicolumn{1}{c|}{\multirow{2}{*}{Method}}  & \multicolumn{2}{c|}{Librispeech} & \multicolumn{2}{c|}{LSM-2mix} & \multicolumn{2}{c}{\begin{tabular}[c]{@{}c@{}}LSM-3mix\\ (zero-shot)\end{tabular}} \\ \cline{3-8} 
                        & \multicolumn{1}{c|}{}                         & Dev             & Test           & Dev           & Test          & Dev                & Test               \\ \hline
S5                      & H-SAGE                                        & 3.6             & \underline{3.8}            & \textbf{6.0}  & \textbf{5.7}  & \textbf{19.7}      & \textbf{19.5}      \\ \hline
S6                      & \quad w/o OA-Loss              & 3.5             & \textbf{3.7}   & \underline{6.1}  & \underline{5.8}  & \underline{20.2}               & \underline{20.1}               \\
S7                      & \quad w/o SA-Encoder + OA-Loss & 3.6             & 4.0            & 6.2           & 6.4           & 20.8               & 21.5               \\ 
S8                      & \quad w/o holistic gating      & \underline{3.4}             & 3.9            & \textbf{6.0}  & 6.2           & 20.4               & 20.9               \\ \hline
S9                      & \quad w MoLE only in FFN       & 3.9             & 4.2            & 6.8           & 6.4           & 21.2               & 21.3               \\
S10                     & \quad w MoLE only in Att.      & \textbf{3.3}    & \textbf{3.7}   & 6.3           & 6.0           & \underline{20.2}               & 20.3               \\ \hline
\end{tabular}%
}
\vspace{-3mm}
\end{table}

\subsection{Architectural Analysis}

As summarized in Table~\ref{tab:res2}, we analyze expert placement across the full H-SAGE (S5), the variant with experts applied only to Feed-Forward Networks (FFN) (S9), and the Attention-only variant (S10). In terms of performance, S5 achieves the best results, followed by S10, with S9 trailing. This ranking confirms that deploying experts in both modules yields a synergistic effect, maximizing the capacity to handle complex acoustic variations.

Notably, our analysis reveals a correlation between global modeling strength and optimal expert placement.
In prior MoE-based MTASR studies, although attention experts are often regarded as crucial, FFN experts can outperform them in certain scenarios due to stronger feature transformation capabilities.
However, the introduction of the SA-Encoder enhances the effectiveness of attention experts.
Our experiments show S10 consistently outperforms S9, indicating that when global features are robustly captured, expert specialization within the attention mechanism becomes pivotal.
Inherently designed for sequence modeling, the attention mechanism is structurally compatible with global cues from the SA-Encoder. This allows it to effectively utilize these cues for speaker disentanglement.
Thus, while FFN experts provide limited support in certain scenarios, attention-based experts become more critical once the acoustic state is effectively modeled.

\subsection{Impact of Auxiliary Loss Weight $\lambda$}

\begin{figure}[t]
    \centering
    \begin{subfigure}[b]{0.48\linewidth}
        \centering
        \includegraphics[width=\linewidth]{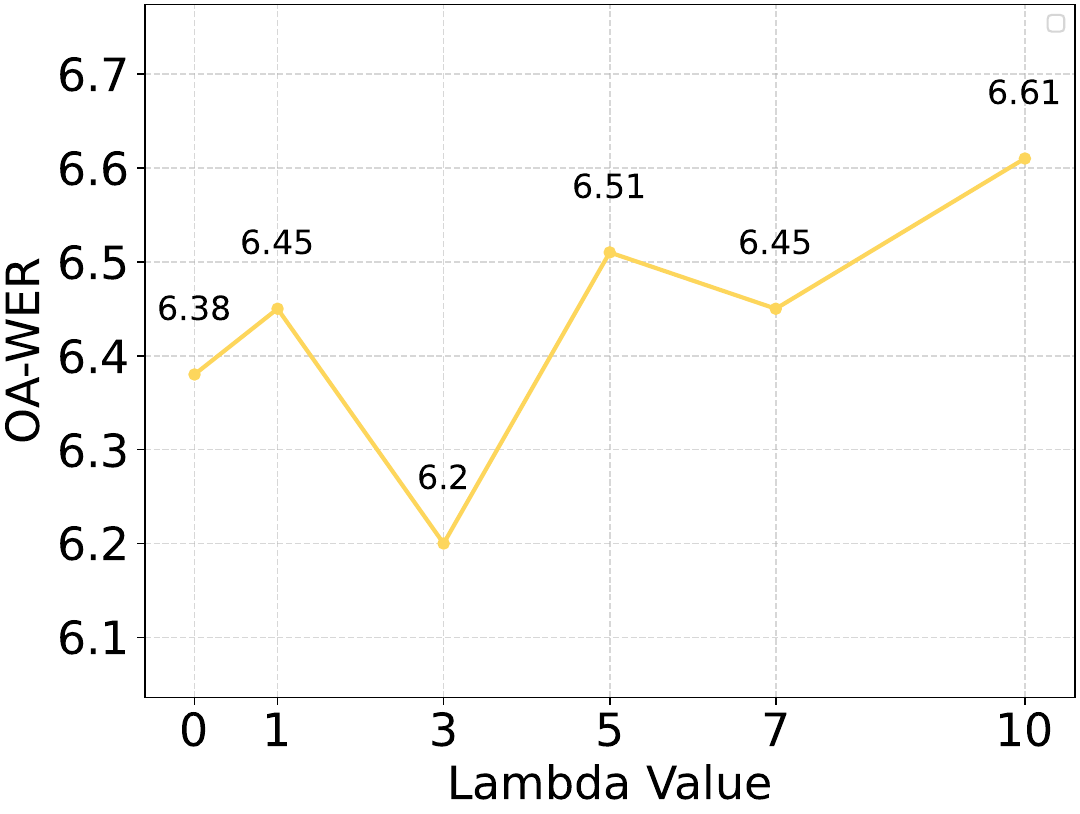}
        \caption{Performance on LSM-2mix}
        \label{fig:left}
    \end{subfigure}
    \hfill
    \begin{subfigure}[b]{0.48\linewidth}
        \centering
        \includegraphics[width=\linewidth]{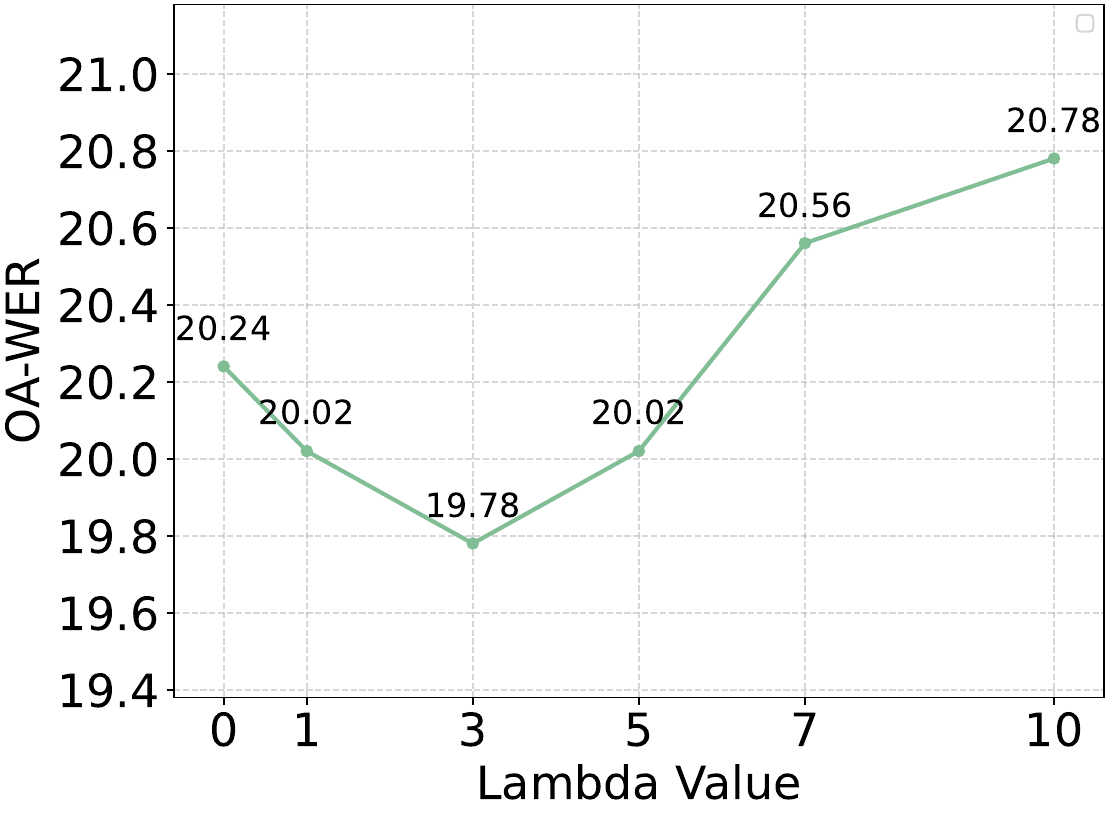}
        \caption{Performance on LSM-3mix}
        \label{fig:right}
    \end{subfigure}
    
    \caption{Impact of the auxiliary loss weight $\lambda$ on OA-WER performance.}
    \label{fig:lab-lambda}
    \vspace{-4mm}
\end{figure}

We evaluate the auxiliary weight $\lambda$ in Eq.~\ref{eq:loss} across the set $\{0, 1, 3, 5, 7, 10\}$. As illustrated in Fig.~\ref{fig:lab-lambda}, we observe that $\lambda=3$ consistently yields optimal performance for both in-domain (LSM-2mix) and zero-shot (LSM-3mix) scenarios.  Notably, while the trend in the simpler 2-mix scenario shows slight fluctuations, the more challenging 3-mix scenario exhibits a distinct U-shaped pattern. This indicates that while the model handles in-domain data relatively easily, explicit supervision of speaker activity states becomes critical for generalizing to complex, unseen acoustic environments. However, further increasing $\lambda$ beyond 3 leads to degradation in both cases, as an overly dominant auxiliary objective overshadows the primary transcription task. Consequently, we set $\lambda=3$ to balance overlap awareness and ASR accuracy.

\section{Conclusion}
In this paper, we propose Holistic Speaker-Aware Guided Experts (H-SAGE), a framework that shifts MoE-based MTASR from implicit adaptation to explicit acoustic modeling. By integrating a Speaker-Aware Global Encoder with a Holistic Gating Mechanism, H-SAGE captures dynamic speaker states and improves expert selection. Experiments on LibriSpeechMix show that H-SAGE outperforms state-of-the-art methods, especially under high-overlap conditions. Furthermore, ablation analyses confirm that explicit acoustic supervision and a holistic view are indispensable for disentangling speakers in complex environments.

\bibliographystyle{IEEEbib}
\bibliography{icme2026references}

\end{document}